# Phase transitions in KNbO$_3$: symmetry analysis and first principles calculations

R.A.Evarestov[a*], Yu.E.Kitaev[b], and S.S.Novikov[a]

[a] St. Petersburg State University, 7/9 Universitetskaya nab., St. Petersburg, 199034, Russian Federation and [b]Ioffe Institute, Politekhnicheskaya 26, St.Petersburg, 194021, Russian Federation.
*Correspondence e-mail: r.evarestov@spbu.ru

## Abstract

Group theoretical aspects of the three temperature-dependent and temperature-reversible experimentally observed phase transitions in the KNbO$_3$ crystal (cubic-tetragonal, tetragonal-orthorhombic, orthorhombic-rhombohedral) in the framework of the group-subgroup relationship tree have been discussed.

The *ab initio* DFT-HSE06 LCAO calculations of the electron and phonon properties, with optimisation of lattice parameters and atomic coordinates for all experimentally observed KNbO$_3$ phases, are used for better understanding of the details of these phase transitions.
 Good agreement with the experimental data was found for the structural properties.
 *Ab initio* calculations of the phonon dispersion curves confirmed the existence of a stable phase only for the rhombohedral structure found experimentally for the lowest temperature of 263 K. For the remaining three higher temperature phases, imaginary frequencies appear, implying a nonstability of these phases.
 The only the cubic-tetragonal phase transition has been found to be symmetry allowed. The tetragonal-orthorhombic and orthorhombic-rhombohedral phases are not related to the group-subgroup relationship. An explanation is proposed based also on the results of ab initio calculations of the structure of the monoclinic phase, which we have chosen as a virtual one for the tetragonal-orthorhombic transition in the bulk.

## 1. Introduction

Symmetry analysis of phase transitions in crystals is usually applied to solve one of two problems (Damnjanović, 1987; Perez-Mato, 2010). The first problem is the prediction of the possible symmetry groups of the low temperature phases, when the symmetry of the high temperature phase is known. The first problem is usually referred to as the Landau problem. The inverse Landau problem is the prediction of higher symmetry phases with the respect to the chosen lower symmetry phase (Kitaev et al, 2015).

In this paper, we study the phase transitions in perovskite potassium niobate ($KNbO_3$) belonging to the $ABO_3$ family ($A$ = K, Ta, … $B$ = Ba, Nb,…). Note, that $KNbO_3$ crystals are isostructural with the $BaTiO_3$ ones having the same sequence of phase transitions (Evarestov, 2011).

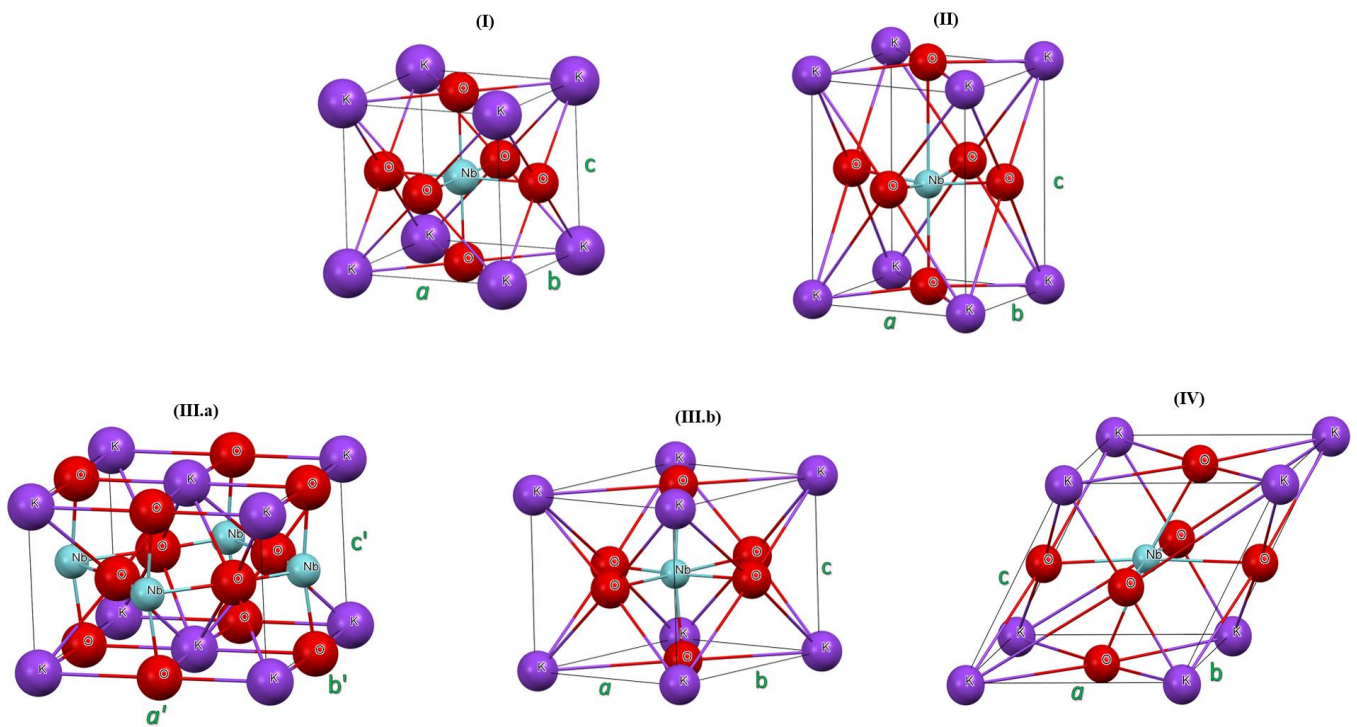

Figure 1: Different phases of $KNbO_3$: (I)Pm-3m, cubic, (II) P4mm, tetragonal, (III) Amm2, orthorhombic (a: conventional orthorhombic unit cell, b: primitive rhombic unit cell), and (IV) R3m, rhombohedral.

Four phases of bulk $KNbO_3$ have been observed experimentally: cubic (Pm-3m, SG 221), tetragonal (P4mm, SG 99), orthorhombic (Amm2, SG 38) and rhombohedral (R3m, SG160) (Skjærvø et al., 2018; Fontana et al 1984).

All these phases are shown in Figure 1. Three consecutive temperature-reversible phase transitions (cubic-tetragonal-orthorhombic- rhombohedral) are detected in the experiment. However the transitions are with hysteresis that is an evidence of their first-order character. Moreover, although the space groups of all the phases are the subgroups of the group of the cubic phase, within the tetragonal-orthorhombic phase sequence their space groups are not connected with the group-subgroup relationship. Thus, the aim of this paper is to elucidate thoroughly this problem by combining the *ab initio* calculations and the group theory analysis using the programs and tools of the Bilbao Crystallographic Server (Aroyo, Perez-Mato *et al*., 2006, 2011 ).

The present paper is organized as follows. In section 2, we discuss group theoretical aspects of the phase transitions in $KNbO_3$. To understand better the details of these phase transitions we use *ab initio* calculations of different properties of $KNbO_3$ depending on the symmetry of the observed phases (Section 3). In Section 4, we discuss the results obtained, and in Section 5 present our conclusions.

## 2. Phase Transitions in $KNbO_3$ crystals: group theory analysis

Four different phases observed in the $KNbO_3$ bulk crystals (Skjærvø *et al*., 2018; Fontana et al 1984), together with the monoclinic phase detected only in nanowires (Kim *et al*., 2013), are shown in Fig. 2. In the boxes corresponding to each phase, the space groups (SG) and the Wyckoff positions occupied by the atoms are given. For temperature-reversible transitions, their temperatures on cooling and heating and the corresponding arrows are shown by blue and red, respectively.

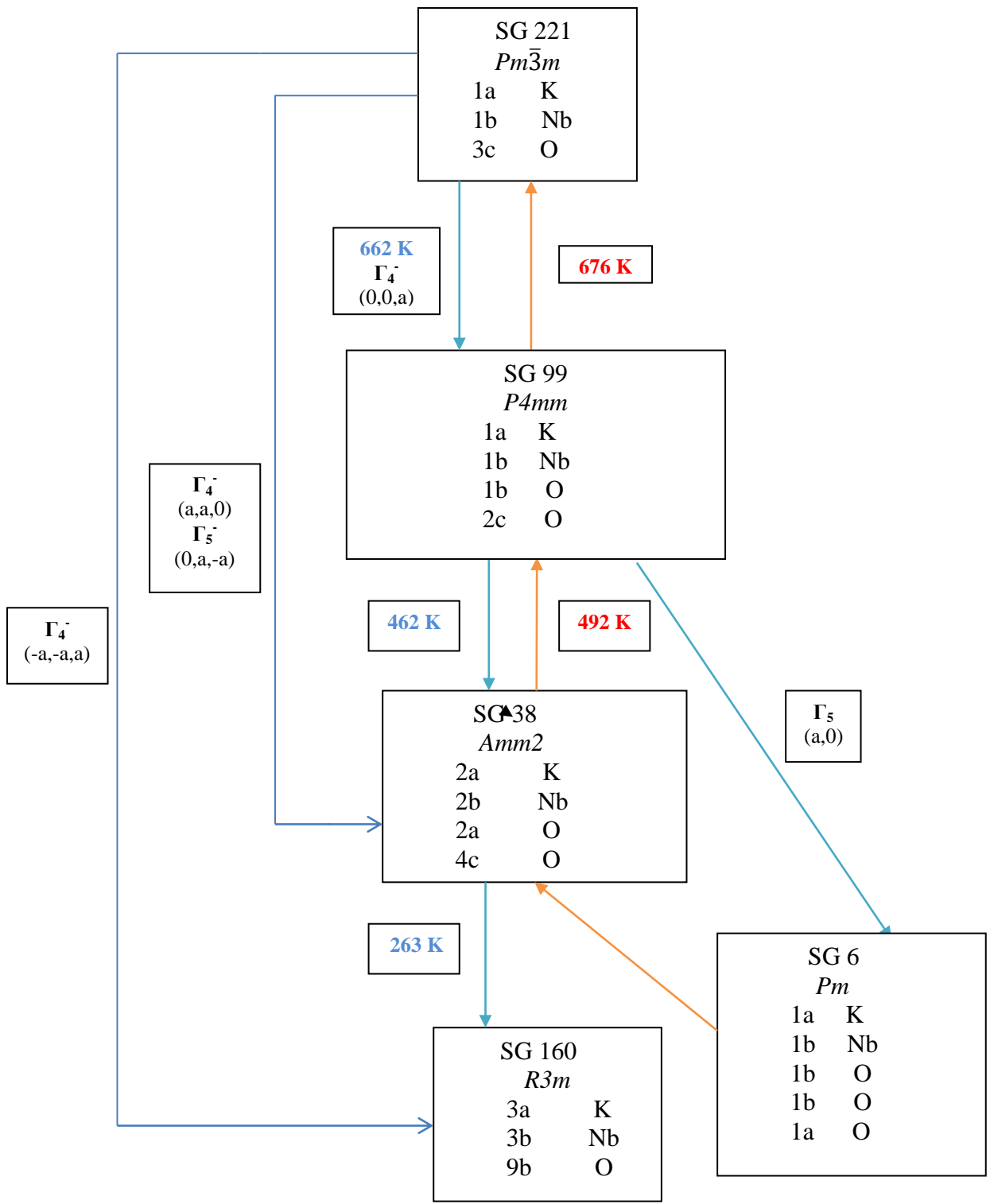

Figure 2. The group-subgroup relationship tree, giving the possible paths between the cubic (SG221 ), tetragonal (SG99), orthorhombic (SG38), rhombohedral (SG160) and monoclinic (SG6) phases. Occupied Wyckoff positions are shown within the corresponding boxes. Transition temperatures on cooling and heating are given near the corresponding arrows shown by blue and red, respectively. Symmetry allowed transitions are marked with active irreps driving the corresponding transitions.

We analyze the transitions between these phases using the programs and tools of the Bilbao Crystallographic Server (Aroyo, Perez-Mato *et al.*, 2006, 2011 ). The AMPLIMODES program (Orobengoa *et al.*, 2009; Perez-Mato *et al.*, 2010) for a pair of the parent structure and the distorted structure allows one to determine the primary and secondary symmetry modes compatible with this phase transition and calculates the amplitudes of the different symmetry-adapted distortions presented in the low-symmetry phase as well as their polarization vectors.

## 2.1. Cubic-tetragonal phase transition

We begin with the cubic phase ($Pm\bar{3}m$, SG 221) observed experimentally at temperatures T > 389 K. The AMPLIMODES program shows that the direct $Pm\bar{3}m$ (SG 221) → $P4mm$ (SG 99) transition into the tetragonal phase is driven by the soft mode $\Gamma_4^-$ with (0,0,a) as the order parameter. The structural data are taken from (ICSD, 2007).

The results of the analysis are shown in Figure 3. From Figure 3, it is seen that $\Gamma_4^-$ mode is the only primary phonon mode which drives the cubic-tetragonal phase transition and no secondary phonon modes participate in the process. The primary mode alone (in contrast to the secondary modes) can lower the symmetry in the phase transition (Hatch & Stokes, 1988).

## Symmetry Modes Summary

| Atoms | WP | Modes |
|-------|----|----|
| O1 | 3c | GM4-(2) |
| Nb1 | 1b | GM4-(1) |
| K1 | 1a | GM4-(1) |

Note: The primary mode is written in bold letters

## Summary of Amplitudes

| K-vector | Irrep | Direction | Isotropy Subgroup | Dimension | Amplitude (Å) |
|----------|-------|-----------|-------------------|-----------|---------------|
| (0,0,0) | GM4- | (0,0,a) | P4mm (99) | 4 | 0.1467 |

**Global distortion: 0.1467 Å**

Figure 3. A screenshot from the AMPLIMODES program for the transition $Pm\bar{3}m \rightarrow P4mm$, showing the only primary active mode $\Gamma_4^-$, the atoms (together with the occupied Wyckoff positions) contributing to this mode, and the amplitude of the distortion connected with the $\Gamma_4^-$ soft mode (normalized with respect to the primitive unit cell of the high-symmetry structure).

The AMPLIMODES program determines also the transformation matrix

```
[ 1 0 0] [0]
[ 0 1 0] [0]
[ 0 0 1] [0]
```

and the Wyckoff position splittings in the tetragonal phase

1a → 1a; 1b → 1b; 3c → 1b+2c

    The tetragonal phase SG 99 is a subgroup of the cubic phase SG 221 whereas the subsequent tetragonal-orthorhombic-rhombohedral phase transitions are not related by the group-subgroup relationship.

## 2.2 Cubic-orthorhombic and cubic – rhombohedral phase transitions

The AMPLIMODE program shows (see Figure 4) that the direct transition from the cubic phase into the orthorhombic (SG 38) one is allowed and driven by two modes $\Gamma_4^-$ and $\Gamma_5^-$.

**Symmetry Modes Summary**

| Atoms | WP | Modes |
|---|---|---|
| O1 | 3c | **GM4-(2)** GM5-(1) |
| Nb1 | 1b | **GM4-(1)** |
| K1 | 1a | **GM4-(1)** |

Note: The primary mode is written in bold letters

**Summary of Amplitudes**

| K-vector | Irrep | Direction | Isotropy Subgroup | Dimension | Amplitude (Å) |
|---|---|---|---|---|---|
| (0,0,0) | GM4- | (a,a,0) | Amm2 (38) | 4 | 0.1840 |
| (0,0,0) | GM5- | (0,a,-a) | Amm2 (38) | 1 | 0.0060 |

**Global distortion: 0.1841 Å**

Figure 4. A screenshot from the AMPLIMODES program for the transition $Pm\bar{3}m \rightarrow Amm2$, showing two primary active modes $\Gamma_4^-$ and $\Gamma_5^-$ being involved into this transition and the amplitudes of the distortion connected with the $\Gamma_4^-$ and soft mode $\Gamma_5^-$ (normalized with respect to the primitive unit cell of the high-symmetry structure).

It is seen that the $\Gamma_4^-$ soft mode distortions involve all the atoms in the unit cell whereas the $\Gamma_5^-$ mode distortion involves only the O1 orbit. Both modes are the primary ones. The soft mode $\Gamma_4^-$ corresponds to (a,a,0) as the order parameter and the soft mode $\Gamma_5^-$ to (0, a, -a) However, as seen from Figure 4 that the amplitudes obtained for the two symmetry-adapted distortions are very different. The amplitude of $\Gamma_5^-$ distortion is more than 30 times smaller.

The SYMMODES program shows that the direct transition from the cubic phase into the rhombohedral (SG 160 ) one is also allowed and driven by the $\Gamma_4^-$ mode with the (-a, - a, a) as the order parameter.

In Section 3, we discuss the results of our *ab initio* calculations of electron and phonon properties of $KNbO_3$.

## 3. *Ab Initio* calculations of $KNbO_3$ electron and phonon properties

### 3.1. Computational details

In the present work, the quantum mechanical calculations of electron and phonon properties for different $KNbO_3$ phases are carried out for two purposes. Primarily, we tried to find the first-principle calculation scheme which correctly reproduces the existing for $KNbO_3$ experimental data and the results of other published calculations. Secondly, the use of the chosen calculation scheme allows better understanding the phase transition details. In particular, the transition between orthorhombic and rhombohedral phases ( see Section 4 ) cannot be explained using group theoretical analysis only.

The computational description of crystal phase stability requires a high accuracy for the calculation of phonon frequencies as they are determined by the second derivatives of the total energy over atomic displacements. By this reason, the high tolerance is needed in direct lattice summations for the overlap threshold in one-electron integrals, for the overlap and the penetration threshold in Coulomb integrals and for overlap threshold in exchange integrals. We took into account this point when choosing the computational details.

We used the hybrid exchange-correlation functional HSE06 (Heyd *et al* , 2003) realized in CRYSTAL17 computer code (Dovesi, Erba *et al* , 2018; Dovesi, Saunders *et al*, 2018). This code is intended for modeling of the periodic systems and localized atomic Gaussian functions are used to expand the Bloch crystal orbitals. Comparison with experimental data for rhombohedral $KNbO_3$ phase results of our calculations with five different density functionals (PBE,PBE0,B3LYP,HSESOL,HSE06) demonstrates the priority of HSE06 hybrid density functional to produce correctly the $KNbO_3$ structure and energy band gap. This conclusion agrees with results of paper (Schmidt *et al*, 2017) where the $KNbO_3$ HSE electronic structure calculations were made after comparison of the results, obtained for five DFT functionals (LDA, PBE, PBEsol, AM05, RTPSS).

The atomic basis sets were taken from the CRYSTAL code site (Dovesi, Erba *et al*, 2018; Dovesi, Saunders *et al*, 2018). All electron basis sets DZVP quality (Consistent Gaussian Basis Sets of Double-Zeta Valence with Polarization quality for solid-state calculations) (Oliveira *et al*, 2019; Peintinger *et al*, 2012) were used for K and O atoms. For Nb atom, the relativistic pseudopotential and the corresponding TZVP ( Triple-Zeta Pseudopotential with Polarization ) quality basis sets for valence electrons (Laun & Bredow, 2022) were applied.

The summation over the Brillouin zone (BZ) was sampled using the 8x8x8 **k**-point Monkhorst-Pack (Monkhorst & Pack, 1976) mesh and the tolerances 8, 8, 8, 8, 16 for the one-electron, Coulomb and exchange integrals were applied. Briefly speaking, these tolerances mean that during the direct lattice summations the one-electron integrals and two-electron Coulomb integrals less than $10^{-8}$ are estimated by the multipolar expansion and two-electron exchange integrals less than $10^{-16}$ are ignored. DFT-D2 approximation (Grimme, 2006) was adopted for the dispersion correction needed for reproducing the van der Waals interactions. The self-consistency in energy with an accuracy of $3 \times 10^{-9}$ eV was achieved when solving the one-electron equations. The geometry of all systems considered was totally optimized until the forces on atoms ware not exceed the value of 0.003 eV/Å

The calculations of phonon frequencies were based on the following procedure ( see application of this procedure for calculations (Evarestov & Bandura, 2012) of phonon frequencies in BaTiO$_3$ four phases). First, the equilibrium geometry was found. In the cubic phase, the lattice parameter, fully defining the structure, was optimized. In the ferroelectric phases, the lattice parameters and fractional displacements of atoms were optimized. Atomic and cell relaxations were performed with the convergence criterion for the forces on atoms set to 0.005 eV/A˚. The threshold on the energy change between optimization steps for the self-consistent cycles was $10^{-8}$ eV for the structure optimization and $10^{-10}$ and for the phonon frequency calculations.

In Section 3.2 we discuss the results of our calculations of the structure and electron properties.

### 3.2. Crystal structure and electron properties

The crystal structures of all the mentioned four KNbO$_3$ phases were determined experimentally and the corresponding data can be found in Inorganic Crystal Structure Database (ICSD, 2007). The outline of the most recent experimental structure data is given in Table 1 ( in parentheses). For each phase, the space group , lattice parameters and occupied by atoms Wyckoff positions are given. Note that for orthorhombic phase Amm2 and rhombohedral phase R3m, the structure data are given for the primitive unit cell ( this setting differs from the conventional setting). Table 1 contains also the comparison of our results on the structure and band gap calculations with those from ( Schmidt F. & Landmann M., *et al.,* 2017) ( marked by*). These results were obtained using a plane-wave implementation of DFT in VASP code (Kresse & Furthmuller, 1996). Non-local HSE hybrid functional was used with the fraction of the exact Hartree-Fock exchange 30 % ( in our HSE06 calculations this fraction was 25%). The electronic wave functions were expanded up to a kinetic energy of 600 eV.

Table 1 demonstrates very good agreement of our calculated data both with results of HSE calculation ( Schmidt *et al.,* 2017) and experimental structure data. In both calculations, the band gap is indirect. In HSE calculations the fraction of Hartree-Fock exchange 30% was taken to match the experimentally observed band gap of the cubic $KNbO_3$ crystal ( in our HSE06 calculations this fraction was 25%).

It is also seen that the calculated total energy decreases in right order with symmetry lowering (cubic, tetragonal, orthorhombic, and rhombohedral) but the band gap value increases when one moves from the high temperature cubic phase to the low temperature rhombohedral phase.

Table 1. Results of atomic structure and band gap calculations with HSE06 Density Functional. The structure data are given in Å. The experimental data are given in brackets and taken from (ICSD, 2007). * mark the results of HSE calculation ( Schmidt *et al.,* 2017)

| Space group (SG) | SG 221 Pm-3m,70* K | SG 99 P4mm, 623 K | SG 38 Amm2, 454.7 K | **SG160 R3m, 343 K** | SG 6 Pm, 273 K |
|---|---|---|---|---|---|
| a | 3.983,3.985* (4.022) | 3.967,3.969* (3.997) | 3.961,3.961* (3.973) | **4.003, 4.001* (4.016)** | 4.024, 4.023* (4.050) |
| b | 3.983, 3.985* (4.022) | 3.967,3.969* (3.997) | 4.024,4.021* (4.035) | **4.003,4.001* (4.016)** | 3.961, 3.961* (3.992) |
| c | 3.983, 3.985* (4.022) | 4.066, 4.058* (4.063) | 4.024,4.021* (4.035) | **4.003, 4.001* (4.016)** | 4.024, 4.020* (4.021) |
| α (α≠90°) | 90° | 90° | 89.84°,89.82* (90.27°) | **89.90, 89.90* (89.82°)** | 90.17, 89.82* (90.10) |
| K | 0 0 0 | 0 0 0 | 0 0 -0.002 (0 0 -0.010) | **-0.002 (3) (0 0 0(3))** | -0.100 0 -0.114 |
| Nb | 0.5 0.5 0.5 | 0.5 0.5 0.517 (0.5 0.5 0.517) | 0.5 0 0.512 (0.5 0 0.516) | **0.510(3) (0.513(3))** | 0.411 0.5 0.372 |
| O1 | 0 0.5 0.5 | 0.5 0.5 -0.027 (0.5 0.5 -0.026) | 0 0 0.481 (0 0 0.481) | **0.5 0.5 0.018 (0.5 0.5 0.018)** | -0.620 0 0.403 |
| O2 | 0.5 0 0.5 | 0.5 0 0.479 (0.5 0 0.480) | 0.5 0.253 0.230 (0.5 0.253 0.232) | **0.5 0.018 0.5 (0.5 0.018 0.5)** | 0.383 0.5 0.095 |
| O3 | 0.5 0.5 0 | 0 0.5 0.479 (0 0.5 0.480) | 0.5 -0.253 0.230 (0.5 -0.253 0.232) | **0.018 0.5 0.5 (0.018 0.5 0.5)** | -0.123 0.5 0.399 |
| Eg (eV) | 2.82, 3.14* (3.14-3.24) | 2.88,3.23* (3.08,3.30) | 3.30,3.59* (3.15,3.25) | **3.53,3.80*** | 4.02,3.63* (3.09,3.15) |
| Delta E (eV) | 0(0) | -0.027 (-0.020) | -0.033 (-0.024) | **-0.034 (-0.026)** | -0.033, ( -0.025) |

Electron energy bands for different KNbO$_3$ phases are shown in Fig. 5

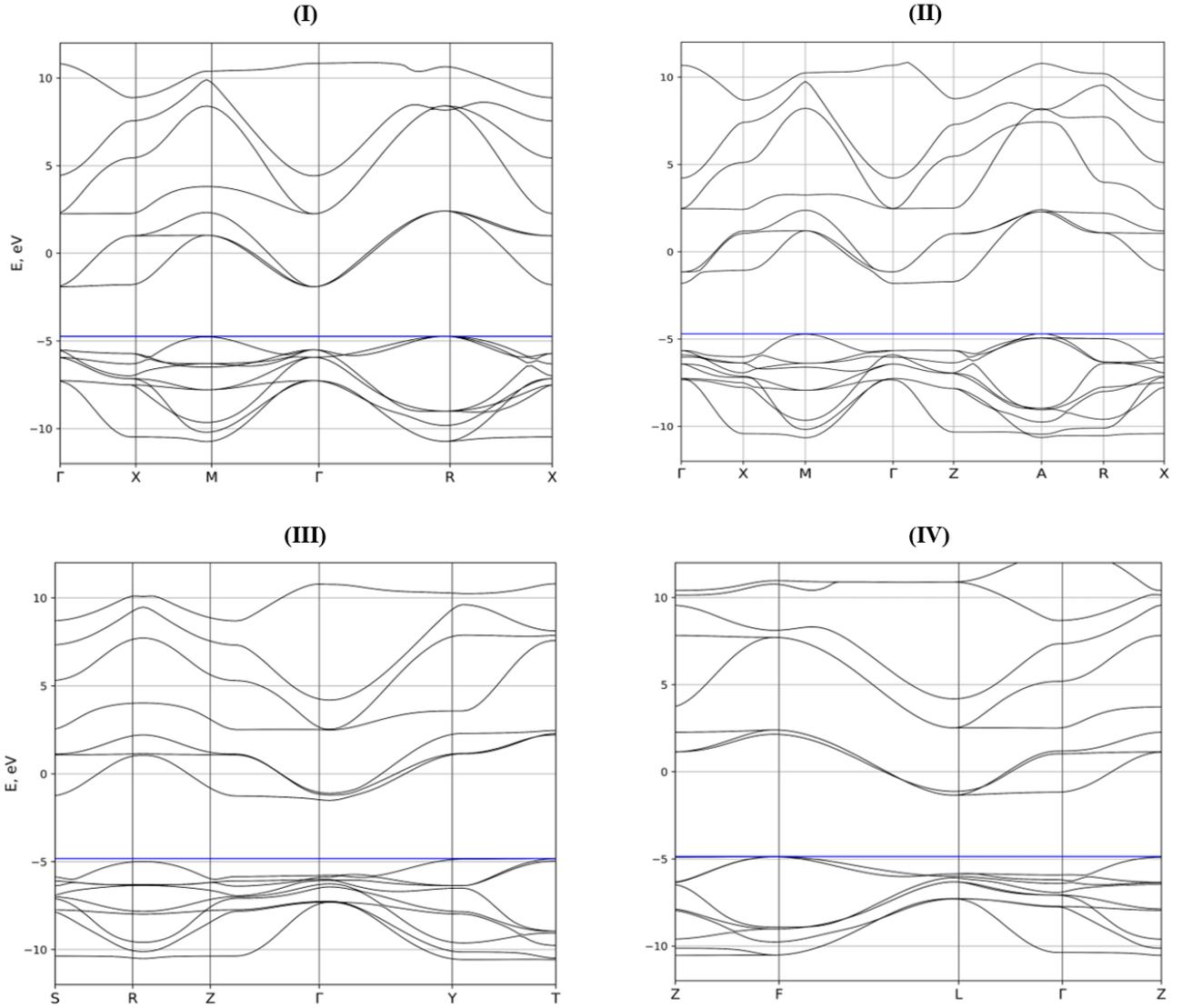

Figure 5. Electron energy bands for different KNbO$_3$ phases: I – cubic phase (SG221) II - tetragonal phase (SG 99) III -orthorhombic phase (SG 38) IV rhombohedral phase (SG 160)

As seen from Figure 5, all band gaps are indirect which agrees with the results of other calculations.

## 3.3. Phonon properties

The phonon frequencies were obtained by the frozen phonon method (Parlinski *et al*, 1997; Evarestov & Losev, 2009) within the harmonic approximation at the optimized equilibrium crystal constants. The zone-center phonon frequencies (the eigenvalues of the dynamical matrix) are determined from numerical second-order derivatives of the ground state energy. To this purpose, the total energy is calculated for found for each crystal phase optimized structure. The supercell approach is used to obtain the phonon frequencies at the nonzero wave vectors. The convergence of the phonon frequencies and dispersion curves in terms of the supercell size was studied in the *ab initio* frozen phonon calculations.

Figure 6 shows phonon dispersion curves for different $KNbO_3$ phases. It is seen that only for rhombohedral phase (SG160) the imaginary phonon frequencies disappear. This confirms the experimental data on the stability of this phase at low temperature.

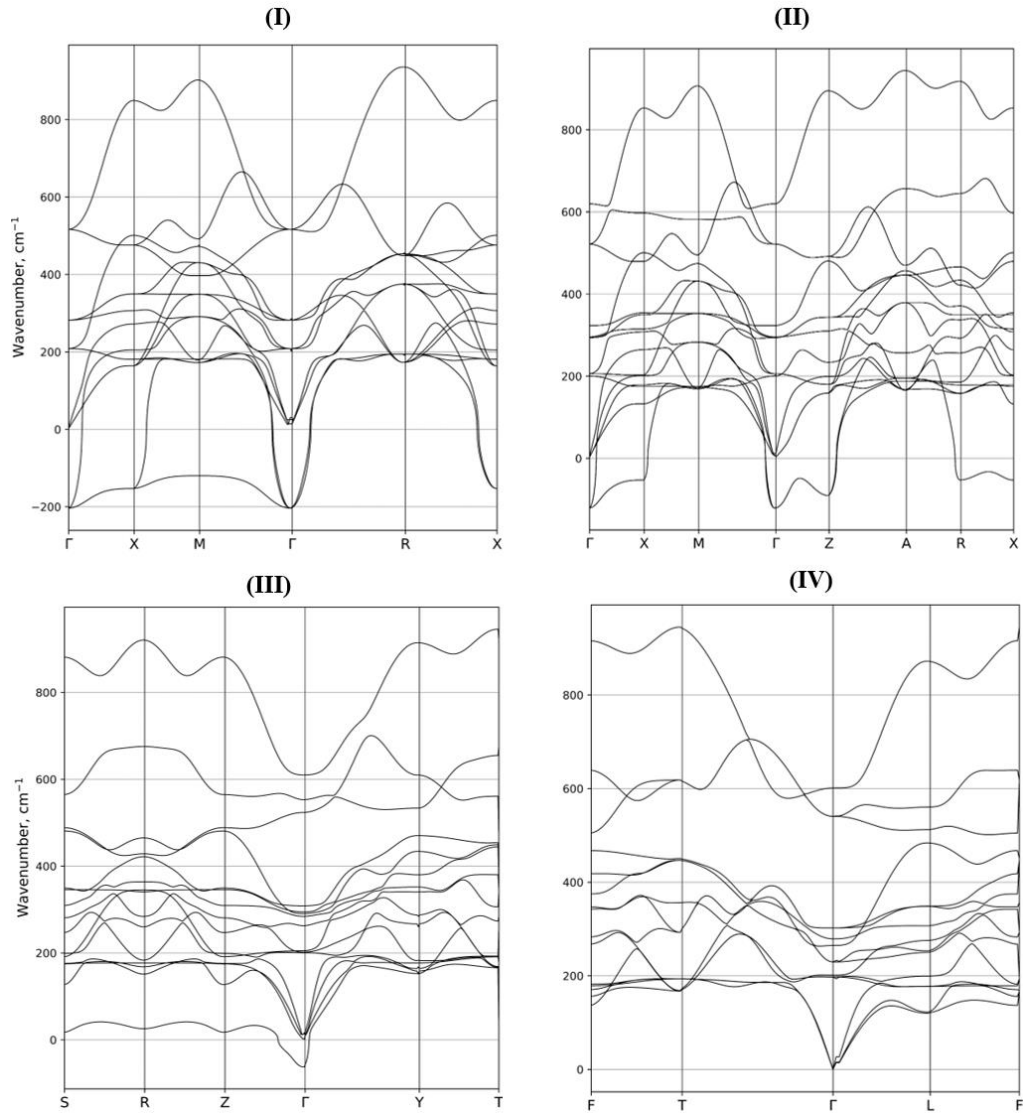

Figure 6. Phonon dispersion for different KNbO$_3$ phases: I – cubic phase (SG221); II -tetragonal phase (SG 99); III -orthorhombic phase (SG 38); IV rhombohedral phase (SG 160)

## 4. Results and Discussions

The temperature reversible sequence of phase transitions has been observed experimentally. *Ab initio* DFT-HSE06 LCAO calculations of the phonon dispersion curves correspond to the temperature 0 K. It confirmed the existence of stable phase only for the lowest temperature rhombohedral structure. For the rest three phases imaginary frequences appear (see Figure 6) meaning a nonstability of these phases for the temperature 0 K.

However, only cubic-tetragonal phase transition is symmetry allowed. The tetragonal-orthorhombic and orthorhombic- rhombohedral phases are not connected with the group-subgroup relationship. The similar transitions in $BaTiO_3$ have been interpreted by (Orobengoa *et al*., 2009) on the base of symmetry-allowed transitions from the cubic phase. A three-fold degenerate polar instability associated with the single active three-dimensional $\Gamma_4^-$ irrep produces three successive ferroelectric phases, by changing its direction within the three-dimensional irrep space.

An alternative explanation could be proposed based also on the results of *ab initio* calculations of the structure of monoclinic phase (SG6), which we have chosen as the virtual one for the bulk. The SYMMODES program gives that the direct transition from the tetragonal phase (SG 99) into the monoclinic (SG6) one is allowed with the $\Gamma_5$ primary mode with (a,0) as the order parameter and with $\Gamma_1$ and $\Gamma_2$ modes as the secondary ones. The comparison of structure parameters of the orthorhombic and monoclinic phases given in Table 1 shows that the monoclinic phase can be considered a good approximation for the orthorhombic one. Therefore, we can consider as the allowed the tetragonal-orthorhombic transition *via* the virtual monoclinic phase.

## 5. Conclusions

*Ab initio* DFT-HSE06 LCAO calculations are made with the optimization of lattice parameters and atomic coordinates for all experimentally observed KNbO3 phases. The good agreement with the experimental data and Plane Wave DFT calculations was found for the electronic and structure properties. The calculated phonon dispersion curves confirmed the existence of stable phase only for the lowest temperature rhombohedral structure. The calculation results perfectly agree with the symmetry analysis.

Skjærvø S.L., Høydalsvik K., Blichfeld, A.B., Einarsrud M.-A. & Grande T. (2018). R.Soc. open.sci. **5,** 180368.